\documentclass[twocolumn,showpacs,preprintnumbers,amsmath,amssymb]{revtex4-1}

\usepackage{graphicx}
\usepackage{dcolumn}
\usepackage{bm}

\begin{document}

\preprint{APS/123-QED}

\title{Evidence for the Formation of Quasi-Bound-State \\in an Asymmetrical
Quantum Point Contact}

\author{Phillip M. Wu\cite{lundnmc}}
\author{Peng Li}
\author{A. M. Chang}
\affiliation{Duke University, Department of Physics, Physics Building, Science Drive, Durham NC 27708}

\date{June 14 2010}

\begin{abstract}
Features below the first conductance plateau in ballistic quantum 
point contacts (QPCs) are often ascribed to electron interaction 
and spin effects within the single mode limit.  In QPCs with a highly 
asymmetric geometry, we observe sharp resonance peaks when the point 
contacts are gated to the single mode regime, and surprisingly, under 
certain gating conditions, a complete destruction of the $2e^2/h$, 
first quantum plateau.  The temperature evolution of the resonances 
suggest non-Fermi liquid behavior, while the overall nonlinear 
characterizations reveal features reminiscent of the 0.7 effect. 
We attribute these unusual behaviors to the formation of a quasi 
bound state, which is stabilized by a momentum-mismatch accentuated 
by asymmetry.
\end{abstract}

\maketitle

The conductance of a quantum point contact (QPC), a narrow constriction 
connecting two regions of an electron gas, is known to be quantized in units of $G_0 = 2e^2/h$ \cite{vanwees1988,wharam1988}. This quantization can be understood within a framework of non-interacting electrons, where the electrons are backscattered by the potential created by the walls of the constriction, and the conductance depends only on the transmission coefficient \cite{landauer1970}. However, additional features below the first quantized plateau at $0.7*G_0$ was observed \cite{thomas1996} and cannot be explained within the Landauer formalism. This 0.7 structure has instead been attributed to electron interaction and spin effects \cite{thomas1998,cronenwett2002,matveev2004,reilly2005,rejec2006,aryanpour2009}. Several scenarios have been proposed to account for the 0.7 structure, including the formation of a quasi-bound state in the constriction \cite{meir2002,hirose2003}, or a novel zigzag Wigner crystal \cite{matveev2004}. Nevertheless, its origin remains controversial.

In this work, we report unusual features in the conductance of QPCs 
with an unconventional, asymmetric gate geometry. In contrast to the 
standard geometry, where two symmetrically placed finger gates define 
the QPC constriction (with no top gate considered), one of the finger gates is replaced by a long wall gate. We find new features, which point to the formation of a quasi-bound state within the QPC constriction: First, sharp resonance peaks are present when the QPCs are gated below $G_0$. The resonance 
line shape conforms to the derivative of the Fermi function, but with an effective temperature which exceeds the lattice temperature at low T. The 
overall behavior strongly suggest non-Fermi liquid behavior---notably, the integrated area under a resonance is not constant, in contrast to a Fermi liquid, and increases as the temperature is increased. The energy spacing between successive peaks also exhibits anomalous behavior, and stays fixed even as their energy position is shifted toward the next sub-band continuum. Second and surprisingly, in some geometries, a complete destruction of the first quantized plateau at $2e^2/h$ is observed. Such a behavior was completely unexpected, and from a theoretical perspective has only been seen in numerical simulations when a quasi-bound state is introduced \cite{bardarson2004,gudmundsson2005}. Other features bear similarity to symmetric QPCs: In particular, nonlinear characterizations reveal features reminiscent of the 0.7 structure, such as the 0.25 plateau at high DC bias, and point to the possibility of ferromagnetic properties. We attribute these striking behaviors to the formation of a quasi-bound state, stabilized by a momentum-mismatch accentuated by asymmetry \cite{song2011}.

\begin{figure}
\includegraphics[width=0.5\textwidth]{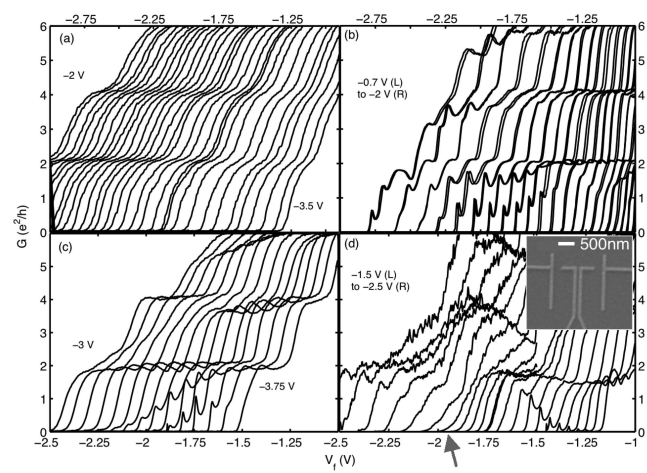}
\caption{\label{fig1} (a) Unequally gated conductance traces in the symmetric geometry: Each trace is obtained with a fixed voltage on one gate (denoted $V_{wall}$) while sweeping the voltage on the other gate ($V_f$). Leftmost trace has $V_{wall}=$-2 V, which is decremented in 50 mV steps until $V_{wall}=$-3.5 V in the rightmost trace. (b)-(d) Traces for asymmetric QPCs with lithographic gap widths of (b) 250 nm (c) 450 nm (d) 300 nm, respectively. The temperature is 300 mK for all traces. Doubling of traces in (b) shows reproducibility. Arrow in (d) indicates a trace with fully suppressed $2e^2/h$ plateau.  Inset shows two QPCs back to back; one is gated shut during measurement.}
\end{figure}

Ti/Au gates were patterned using electron beam lithography on 
a GaAs/AlGaAs heterostructure containing a 2-dimensional electron 
gas 80 nm below the surface, with carrier mobility 
$\mu=9 \times 10^5$ cm$^2$/Vs and density 
$n_{2D} = 3.8 \times 10^{11}$ cm$^{-2}$. Voltages were applied 
unequally to the QPC gates by fixing one while sweeping the other.  
Conductance traces were obtained by applying an excitation voltage of 
$V_{AC}=10 \mu$V ($< k_BT/e$, where $k_B$ is the Boltzmann constant) 
across the QPC at 17.3 Hz, and measuring the current using a lock-in 
amplifier after conversion to a voltage in an Ithaco 1211 current amplifier.

In Fig. \ref{fig1}, we contrast the behavior in a standard geometry with symmetric gates (panel (a)), with those in the asymmetric 
geometry of differing gaps (panels (b)-(d)). In the symmetric case, one finger gate (labeled $V_{wall}$) is held fixed, while the other (labeled $V_f$) is swept. The unequal voltage gating shifts the conductance plateaus as $V_{wall}$ is stepped. Typical of several devices made, the 0.7 structure is observable in several traces between $V_f =$ -2.5 V to -1.5 V; no other anomaly is found regardless of the gap width \cite{wakaya1996}. In the series of traces at an increasingly negative fixed $V_{wall}$ voltage (left to right), the confinement potential across the gap is expected to become sharper \cite{buttiker1990}, as the total number of quantized plateaus observed decreases.  At the same time, the pinchoff voltage, where the conductance is shut to zero, shifts rightwards, while the density in the single channel (mode) limit decreases.

\begin{figure}
\includegraphics[width=0.5\textwidth]{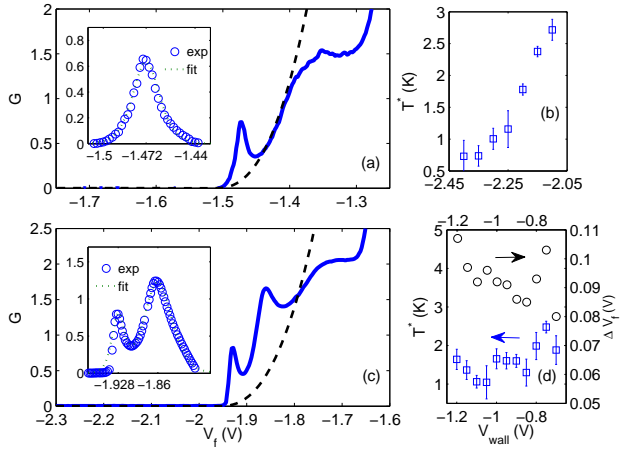}
\caption{\label{fig2} (a) Expanded conductance versus $V_f$ plot for asymmetric geometry with 300 nm gap, at $V_{wall}=$-2.25 V, showing a resonance peak. (b) Extracted effective temperature $T^*$ as a function of $V_{wall}$. (c) Expanded plot for asymmetric geometry with 250 nm gap, at $V_{wall}=$-1 V. (d) Extracted temperature $T^*$ for the left peak and double-peak spacing $\Delta V_f$ as a function of $V_{wall}$. The right peak exhibits a similar behavior for $T^*$. Insets in (a),(c) show the resonances after background subtraction. The resonance peaks can be fit to the derivative of the Fermi function as shown.  The temperature is 300 mK.}
\end{figure}

The behaviors in the asymmetric devices are notably different as shown in Figs. \ref{fig1}(b)-(d). Numerous features are present below $G_0$. Features are reproducible upon multiple voltage scans within the same cooldown, as shown in (b). Structure is also present at higher conductance, but here, we focus on the single channel (mode) limit. In particular, sharp resonances below $G_0$ are observed for all QPC gaps. In what follows, these peaks will be shown to exhibit characteristics consistent with resonant tunneling in a non-Fermi liquid, and specifically, a Luttinger liquid. Figure \ref{fig2}(a) shows an expanded view of a resonance for the 300 nm gap sample at $V_{wall} = -2.25 V$.  The resonance resides on a conductance background and is re-displayed in panel (b) after background subtraction. A background of the form $constant*((V_f-V_p)/x)^y$ was assumed, with $V_p$ the pinchoff voltage.  
The lineshapes can be fit to the derivative of the Fermi function, $G=(e^2/h)*\Gamma*(\pi/(4k_BT^*))*(cosh(\alpha*(V_f-V_{pk})/(2k_BT^*)))^{-2}$. Here $\alpha$ is the lever-arm parameter, $V_{pk}$ the voltage at a resonance peak, and $\Gamma$ is the intrinsic width of the resonance. The effective temperature $T^*$ exceeds the lattice temperature, shown in panel (b), and lifetime broadening of the peaks dominates rather than thermal broadening. The subtraction of a background is predicated on the idea that even in the absence of the resonance, there is a non-zero background transmission in making the transition from zero to the first plateau.  Qualitatively similar behavior is found by subtracting a linear background instead.

\begin{figure}
\includegraphics[width=0.5\textwidth]{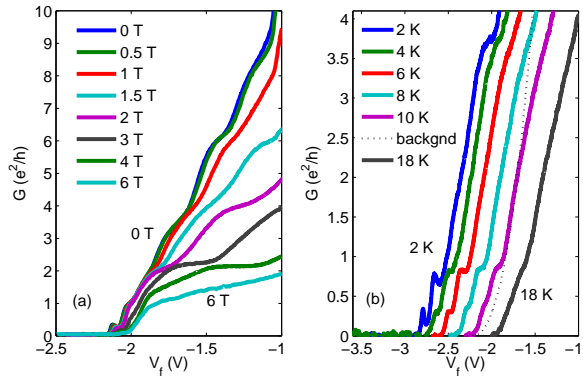}
\caption{\label{fig4} (a) Evolution of the suppressed $2e^2/h$ plateau in an increasing external magnetic field applied perpendicular to the 2DEG plane at 4.2 K. The $2e^2/h$ plateau recovers around or below 1 T, suggesting an energy of $0.43 - 0.85 meV$ ($5\sim10$ K) (taking the value at mid-gap for the Landau gap of 1.7 meV at 1 T). (b) Temperature dependence of the suppression--traces offset 
horizontally for clarity. The $4e^2/h$ plateau is thermally smeared by 6 K, but the resonance features near $e^2/h$ and $(0.5)e^2/h$ persist up to 10 K, consistent with the energy-scale set by the field dependence.  Even at 18 K, an inflection in the conductance at $e^2/h$ is still discernible. Dashed line shows the subtracted background for Fig. \ref{fig5}(a). Note that the slope just above the resonance features appears to be independent of temperature for traces between 2 K - 10 K.}
\end{figure}

We estimate the lever-arm $\alpha$, which relates $V_f$ measured from pinch-off to an energy, from magnetic subband depopulation in a perpendicular magnetic field \cite{vanwees1988prb,wharam1989,weisz1989}. For example, at a typical setting of $V_{wall} \sim -1.6 V$ in Fig. \ref{fig4}(a), we find the Fermi energy $E_F$ to be approximately linear in $V_f$. At the center of the first conductance plateau, $E_F$ has a value $5 - 7$ meV, depending on whether a harmonic oscillator or a square well is used to model the lateral confinement potential. This value is expected to increase as the confinement becomes sharper. For the 300 nm gap device, using the lower bound of $E_F \sim 5$ meV and the difference $\Delta V_f \approx 200$ mV between threshold and the center of the conductance plateau, we find a value of $\alpha = 1/40$. This allows for the effective temperature width to be extracted, which is found to be nearly constant over the range $-2.45$ V$ \le V_{wall} \le -2.25$ V. On the other hand, the background conductance at the peak position has increased by a factor of $\sim 5$. In a Fermi liquid, typically the background conductance is roughly proportional to the inverse life time of the resonance, as is the width. Thus they are expected to scale together. In our devices, this scaling is grossly violated, indicating unconventional behavior. Similar behavior for a double-peak is shown for the 250 nm gap device in Fig. \ref{fig2}(c).

A second unusual signature is found in the energy spacing between 
resonances, measured in the difference in their $V_f$ positions, $\Delta V_f$.  
The simplest interpretation of a double resonance is that the quasi-bound state radius is relatively large, allowing two energy levels to exist below the second subband continuum. In Fig. \ref{fig2}(d), we show the energy spacing $\Delta V_f$ for the 250 nm gap device of Fig. ~\ref{fig1}(b). Surprisingly, the spacing is independent of $V_{wall}$. The position of the double-peak relative to pinchoff $V_f$ varies monotonically as $V_{wall}$ is varied.  Thus the energy position shifts correspondingly. For a Fermi-liquid, this spacing tends to decrease as the states come closer in energy position to the continuum (second subband). Here, this does not occur, and the spacing is nearly constant. This highly unexpected behavior is found in all devices exhibiting a double-peak, including those of Fig. ~\ref{fig1}(b) and (c), and three additional devices as well.

\begin{figure}\includegraphics[width=0.5\textwidth]{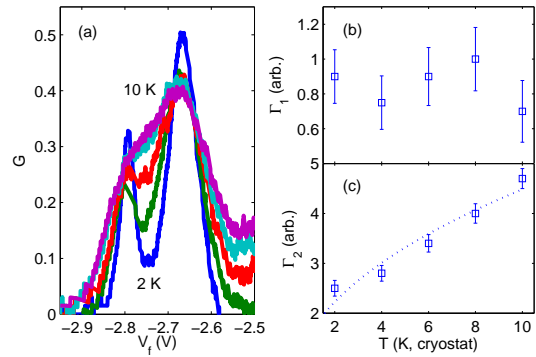}
\caption{\label{fig5} (a) Data of Fig. \ref{fig4} for T = 2, 4.2, 6, 8, and 10 K, with a temperature independent background subtracted, to reveal the temperature evolution of the double-peaks. (b) Extracted $\Gamma$, which is proportional to the integrated area under a resonance, for the left peak ($\Gamma_1$), versus temperature. (c) $\Gamma$ for the right peak ($\Gamma_2$)). Dotted line a fit to $T^{1/g-1}$ with $g \approx 0.7$.}
\end{figure}

A third indication is found in the temperature evolution of the resonances. In Fig. \ref{fig5}, we present data on a second 300 nm gap device and study the integrated area under the resonances for temperatures between $2 - 10 K$. After subtraction of a temperature independent background (dashed line in Fig. 3(b)), the double-peak is again fitted to the derivative of the Fermi function. A temperature independent background subtraction was employed instead of a rising background with rising temperature as the latter yielded widths below the thermal-broadened width, and is thus non-sensical. The area is proportional to the product of the extracted height and width. A hallmark signature of a Fermi liquid resonance is that the area is preserved; in other words, the product should be independent of 
temperature. In fact, as would be expected for a Luttinger liquid 
resonance, the area increases as the temperature is raised. In Fig. 
\ref{fig5}(b),(c), we plot the product $\Gamma$ for both peaks. The area under the left peak is roughly 3 times smaller than that of the right peak. For this peak the uncertainty is too large to arrive at any systematics. For the right peak, however, the product is found to scale roughly as $T^{0.4 \pm 0.1}$. The corresponding Luttinger parameter $g \sim 0.7$\cite{kane1992prl,kane1992prb,auslaender2002}. This is higher than the background conductance ranging between $0.1 - 0.55 e^2/h$ for the peak but similar to what was found in the literature \cite{auslaender2002}. The discrepancy may arise from the finite and short length of the 1D channel, and thus the Luttinger liquid segment.

Aside from the resonances, surprisingly, for the 300 nm 
gap, the $2e^2/h$ plateau is found to be completely suppressed under 
certain gating conditions. This suppression is also found in a 
250 nm gap device, while the general weakening of the plateau 
(less flat) can be seen in all devices. For some devices, this weakening occurs quasi-periodically in $V_{wall}$ settings. The behavior found in our 300 nm device of a direct jump in conductance from pinchoff to $4e^2/h$ has been suggested to be related to formation of incipient lattice in the case of a symmetric QPC \cite{hew2009,smith2009,sfigakis2008}. Alternatively, numerical simulations have also shown such features to be possible when considering an impurity embedded in the 1D channel, when the impurity potential is attractive and gives rise to a quasi-bound state within a self-consistent model \cite{bardarson2004,gudmundsson2005}. The stabilization of a quasi-bound state have been proposed to arise from Coulomb interaction \cite{meir2002,hirose2003}. Thus, the suppression may arise from a high energy scale Coulomb interaction in the QPC bound state, although a recent theoretical calculation suggests that momentum-mismatch alone is sufficient to give rise to a quasi-bound state \cite{song2011}. Whereas in symmetric QPCs the existence of a quasi-bound-state is strongly debated \cite{meir2002,sfigakis2008}, and detailed studies appears to rule out their presence, in the asymmetric QPCs studied here, the momentum-mismatch could be accentuated, and a quasi-bound state may be more likely to form.

An estimate of the energy scale of the suppression 
is obtained by applying a magnetic field $B_z$ perpendicular
to the 2DEG plane, as shown in  
Fig. \ref{fig4}(a). No change is observed up to 0.5 T, which provides 
a lower bound estimate of $\sim 0.43 meV$ ($\sim 5 K$) (taking the value at half the 
Landau gap). At 1 T, an inflection near $2e^2/h$ suggests an upper 
bound of 10 K. This is a rather large energy scale. Beyond 3 T, the 
system enters the filling factor $\nu=2$ quantum Hall regime, as only 
the ground subband remains, and a $2e^2/h$ plateau is clearly 
visible, indicating cyclotron confinement has exceeded Coulomb energy. At stronger fields, signature of the Zeeman spin gap at $e^2/h$ 
re-develops.  In a magnetic field applied {\it parallel} to the 
the 2DEG of up to $6 T$ (data not shown), no change is observed, 
suggesting the spins may already be polarized at zero field.  
Temperature dependence of the suppressed 
$2e^2/h$ plateau is shown in Fig. \ref{fig4}(b). The $4e^2/h$ plateau 
is thermally smeared out above 6 K; however, 
features below $\sim e^2/h$ persist up to 10 K. Even at 18 K, 
the features near pinch-off are smeared, but an inflection in the 
curvature near $e^2/h$ remains, again supporting the high energy scale 
estimate of electron interactions.

\begin{figure}
\includegraphics[width=0.5\textwidth]{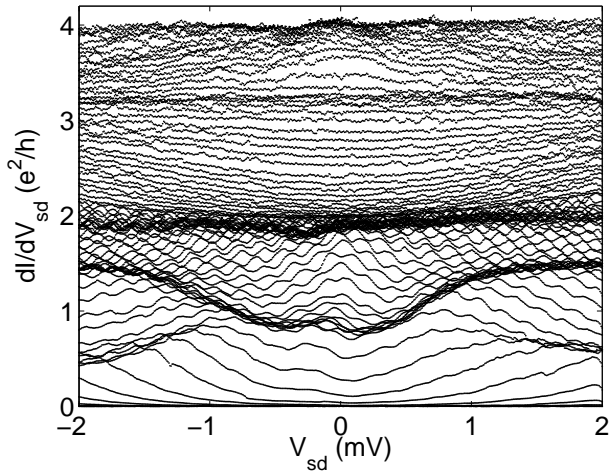}
\caption{\label{fig6} DC source drain bias dependence ($V_{SD}$) for the 250 nm gap sample at zero applied magnetic field and $V_{wall}=$-1.2 V. Strong bunching at $e^2/h$ and transitions to $1.5e^2/h$ at $\sim$1 mV are seen. All traces measured at 300 mK in an Oxford He3 refrigerator.}
\end{figure}

In Fig. \ref{fig6}, the differential conductance versus DC 
source-drain bias at 300 mK is presented, obtained by adding an AC bias to 
the DC bias and measuring the AC signal in a lock-in amplifier. The bunching 
of the traces in the differential conductance versus DC bias corresponds to quantization plateaus in the nonlinear differential conductance. Note that an $e^2/h$ plateau typically develops under a strong magnetic field which spin polarizes the electrons. Here, no magnetic field is applied.  
At a high DC bias around 1 mV, the $e^2/h$ bunch evolves to $1.5e^2/h$, 
which is reminiscent of the 0.7 effect \cite{thomas1996}.  Based on this 
voltage, we estimate the spin splitting to correspond to an applied field of nearly 40 T \cite{cronenwett2002}. Formation of a ground state magnetic moment is often surmised in light 
of the strong electron interaction effects in our 
QPC \cite{rejec2006}. More specifically, Hund rule coupling of the 
electrons strongly bounded to the QPC can give rise to the magnetic 
moment \cite{song2011}. The additional ripples in the differential conductance 
likely arise from interference effects \cite{liang2001}, and are 
currently under investigation. 

In conclusion, we present evidence for formation of a quasi-bound 
state in an asymmetrical QPC. First, resonant peaks are observed for all 
devices of different gap widths. The multiple resonances are unique to our geometry, and provide direct evidence for a quasi-bound state. The peaks have a line-width independent of the background conductance and an area that exhibits a power law dependence on temperature, suggesting non-Fermi liquid behavior. The quasi-bound state is likely stabilized from an exaggerated  momentum-mismatch at the openings of the asymmetric geometry. Second, the unusual suppression of the $2*e^2/h$ plateau also suggests a bound state \cite{song2011}, and possibly strong interaction effects \cite{hew2009}. Finally, nonlinear differential conductance data indicate possible formation of magnetic moment in the bound state, and many features are reminiscent of the phenomenology associated with the 0.7 effect in symmetric devices.  Our findings suggest that the quasi-bound state and interaction effects combine to give rise to the unusual behaviors observed.


\begin{acknowledgments}
We would like to thank Harold Baranger, Eduardo Mucciolo, Gleb Finkelstein, and Maw-kuen Wu for useful discussions, Stephen Teitsworth for use of his dewar and Mike Melloch for crystal. This work was supported in part by the Institute of Physics, Academia Sinica, Taiwan, and by NSF DMR-0701948.\end{acknowledgments}

\bibliographystyle{apsrev}

\end{document}